\documentstyle[a4]{article}

\begin{document}

\newcommand{\bib}{\bibitem}
\newcommand{\er}{\end{eqnarray}}
\newcommand{\br}{\begin{eqnarray}}
\newcommand{\be}{\begin{equation}}
\newcommand{\ee}{\end{equation}}
\newcommand{\epe}{\end{equation}}
\newcommand{\bea}{\begin{eqnarray}}
\newcommand{\eea}{\end{eqnarray}}
\newcommand{\ba}{\begin{eqnarray}}
\newcommand{\ea}{\end{eqnarray}}
\newcommand{\epa}{\end{eqnarray}}
\newcommand{\ar}{\rightarrow}

\def\r{\rho}
\def\D{\Delta}
\def\R{I\!\!R}
\def\l{\lambda}
\def\D{\Delta}
\def\d{\delta}
\def\T{\tilde{T}}
\def\k{\kappa}
\def\t{\tau}
\def\f{\phi}
\def\p{\psi}
\def\z{\zeta}
\def\ep{\epsilon}
\def\hx{\widehat{\xi}}
\def\na{\nabla}
\begin{center}

{\bf Parent Action Approach for the Duality between Non-Abelian
  Self-Dual and Yang-Mills-Chern-Simons Models.}

\vspace{1.3cm} M. Botta Cantcheff\footnote{e-mail: botta@cbpf.br}

\vspace{3mm} Centro Brasileiro de Pesquisas Fisicas (CBPF)

Departamento de Teoria de Campos e Particulas (DCP)

Rua Dr. Xavier Sigaud, 150 - Urca

22290-180 - Rio de Janeiro - RJ - Brasil.
\end{center}

\begin{abstract}

 It has been argued by some authors that the parent
 action approach cannot be used in order to establish the duality
 between the 2+1 Abelian and non-Abelian Self-Dual (SD) and
 Yang-Mills-Chern-Simons (YMCS) models for all the coupling regimes.
 We propose here an alternative (perturbative) point of view, and show that this
 equivalence can be achieved with the parent action approach.

\end{abstract}

\vspace{5mm}

There is a well-known duality between the (2+1)-dimensional Maxwell-Chern-Simons
and Self-Dual \cite{DJ} Abelian models; one can construct a
 so-called parent action \cite{suecos} to show this result
\cite{DJ,KLRvN,BFMS}. Here, we propose this same issue but viewed in an alternative way, which
 allows the generalization to the non-Abelian case. This is the main
contribution of this letter.

 The non-Abelian (NA) version of the so-called Self-Dual model \cite{TPvN}
 presents some well-known dificulties in order to establish the dual
  equivalence to the YMCS theory \cite{BFMS} for the full range of the coupling
 constant.
 The parent action approach first proposed by Deser and Jakiw \cite{DJ} has proven to
 be useful
  in exhibiting the dual equivalence in the Abelian case; however,
 the situation is less understood in the non-Abelian case, where this
 equivalence has only
  been set up for the weak coupling regime \cite{BFMS}. In \cite{KLRvN,BBG} it is
 argued that the
  use of parent actions in this situation is ineffective to establish
 this duality since YMCS (or reciprocally SD) 
  results to be dual to a non-local theory.

 Recently, a technique claimed to be \cite{AINRW} alternative to the parent action
 approach
 has been shown to give the expected result for the Abelian case; then, it
is inferred to
  work also for the non-Abelian case and for other cases too \cite{AINRW,clI,clI2}.
 This method is
 based on the traditional idea of a local lifting of a global symmetry and
 may be realized
 by an iterative embedding of Noether counterterms. However, whenever applied to the
 non-Abelian case \cite{clI},
this method does not provides with a proof, but rather
 a suggestion for this equivalence. The main purpose of this letter is to set a
direct proof based on the parent action approach.

  This is (briefly) the updated
  scenario for this problem. In this work, we proceed further
  and propose a novel way to solve the difficulties with the
  parent
  action suggested in Ref. \cite{DJ}, based on a perturbative analysis,
  and manifestly show the dual
  correspondence between the
  non-Abelian SD and YMCS models for the full range of the coupling constant,
  extending the proof proposed by Deser and Jackiw in the Abelian domain.

 We shall show here that the
  parent action proposed in Ref. \cite{DJ}
actually interpolates YMCS with a (dual) theory, whose action is
SD up to {\it fourth} order in the field.

 The so-called Self-Dual Model \cite{DJ,TPvN,annals} is given by
the following action,
\be
\label{180}
 S_{SD}[f]= \int\, d^3x\:\Bigg(\frac{\chi }{2}\,
\epsilon_{\mu\nu\lambda}\,f^\mu\,\partial^\nu f^\lambda +
\frac{m}{2}\, f_\mu f^\mu \Bigg)\,. \ee

 This actually is a
self-dual model since the duality operation is, in
2+1-dimensions
 commonly defined by
\be \label{210} {}^{\star}f_\mu = {\chi \over m}\,
\epsilon_{\mu\nu\lambda}\,
                  \partial^\nu f^\lambda\,,
\ee
Self or anti-self duality is
 dictated by $\chi = \pm 1$ and $m$ is a constant to render the
  ${}^{\star}$-operation dimensionless.
 Here the Lorentz
indices are represented by greek letters taking their usual values
as $\mu , \nu , \lambda = 0,1,2$. The gauge invariant combination
of a Chern-Simons term with a Maxwell action \be \label{370}
S_{MCS}[A]= \int\, d^3x\:\Bigg(\frac{1}{4m^{2}}
F^{\mu\nu}F_{\mu\nu} -
\frac{\chi}{2m}\,\epsilon^{\mu\nu\lambda}\,A_{\mu}\,
\partial_{\nu}A_{\lambda}\Bigg)\,,
\ee
is the topologically massive theory, which is known to be equivalent \cite{DJ} to
 the self-dual model (\ref{180}).
$F_{\mu\nu}$ is the usual Maxwell field strength,

\be \label{285} F_{\mu\nu}[A] \equiv \partial_{\mu}A_{\nu} -
\partial_{\nu}A_{\mu}\, = 2\partial_{[\mu}A_{\nu]}. \ee

This equivalence has been verified with the parent action approach \cite{DJ,FS}.
 We propose here an alternative way to generalizate it
 to the Non-Abelian case.

The non-Abelian version of the vector self-dual model (\ref{180}),
which is our main concern in this work, is given by

\be  
\label{380} 
{\cal S}_{SD}[f]\equiv \int\,d^{3}x~ \frac{\chi }{2}\,
\epsilon^{\mu\nu\lambda}\,
 \left({f}_\mu^{~a}\partial^{}_{\nu}f_{\lambda}^{~a} +
\,\frac{\tau^{abc}}{3}\,{ f}_{\mu}^{~a} \,{ f}_{\nu}^{~b} 
\,{f}_{\lambda}^{~c}\right) - \frac{m}{2}\, f_{\mu}^{~a} f^{\mu a },
 \ee

 where ${f}_{\mu} =
f_{\mu}^{a}{\tau}^{a}$ is a vector field taking values in the Lie
algebra of a symmetry group $G$ and ${\tau}^{a}$ are the matrices
representing the underlying non-Abelian gauge group with $a=
1,\ldots , \mbox{dim}\:G$; $\tau_{abc}$ are the structure
constants of the group \footnote{We assume that $f_\mu$ is in the adjoint
 representation.}. 

The field-strength tensor is now defined
as \be { F}_{\mu\nu}[A] =
\partial_{\mu}{A}_{\nu} -
\partial_{\nu}{A}_{\mu} + [{A}_{\mu}\,,{
A}_{\nu}] \,, \label{390} \ee and the covariant derivative is
${ D}_{\mu} = \partial_{\mu} + [A_{\mu}\,,\, ]$,
where $A_\mu$ is also a vector field in the adjoint representation
of the group $G$. This may be written 
with explicit group indices, using 
$\left[{\tau}^a,{\tau}^b\right] = \tau^{abc}\tau^{c}$; the field-strength reads as
\be
 { F}_{\mu\nu}^{~~a}[A] =
\partial_{\mu}{A}_{\nu}^{~a} -
\partial_{\nu}{A}_{\mu}^{~a}   + \tau^{abc}{A}_{\mu}^{~~b}~{
A}_{\nu}^{~~c} \,. \label{391} \ee 

Using the Parent action approach, the action (\ref{380}) has been shown (in Ref. \cite{BFMS})
to be equivalent to the gauge invariant Yang-Mills-Chern-Simons (YMCS) theory
\be {\cal S}_{YMCS}[A] = \int\,d^{3}x\,tr\left[\,
\frac{1}{2m}\,{ F}^{\mu\nu a}[A]\,{ F}_{\mu\nu}^{~~a}[A]  -
\chi\,\epsilon^{\mu\nu\lambda}\left({A}_{\lambda}^{~a} \,\partial^{}_\mu
A^{~a}_\nu   + \frac{\tau^{abc}}{3}\,{ A}_{\mu}^{~a} \,{ A}_{\nu}^{~b} 
\,{A}_{\lambda}^{~c}  \right) \,\right]\,, \label{510} \ee only in the
weak coupling limit $m^{-1}\to 0$ so that the Yang-Mills term
effectively vanishes \footnote{The coupling constant is given by the mass parameter,
through $g^2 \equiv \frac{4\pi}{m}$}. In order to establish the dual equivalence
between (\ref{380}) and (\ref{510}) for all
 coupling regimes, we write down the general
 parent action, which clearly contains the one for the Abelian case \cite{DJ}:

 \be
 \label{masterDJ}
 {\cal S}_{Parent}[A,f]= {\cal S}_{CS}[A] -\int\,d^{3}x\, \left[ \epsilon^{\mu\nu\lambda}\,
F_{\nu\lambda}^{~~a} [A]
 f_{\mu}^{~a} \, +  m f_{\mu }^{~a} f^{\mu a}\right] ,
\ee

where 
\be
\label{CS}
{\cal S}_{CS}[A] \equiv \int\,d^{3}x\, \epsilon^{\mu\nu\lambda}\, \left( 
{A}_\mu^{~ a}\partial^{}_{\nu}A_{\lambda}^{~a} + \frac{\tau^{abc}}{3}\,{ A}_{\mu}^{~a}
 \,{ A}_{\nu}^{~b} \,{A}_{\lambda}^{~c} \right).
\ee

First, we shall observe that the master action is invariant
 front the transformations; $ A_{\mu} \to \Delta^{-1} A_{\mu} \Delta +
 \Delta_{\mu}$,  $~~f_\mu \to \Delta^{-1} f_\mu \Delta$.
 Where $\Delta_{\mu}$ is a {\it pure gauge}: $\Delta_{\mu}:=\Delta^{-1} \partial_{\mu}
 \Delta$, and $\Delta$ is a group element.
 We can verify this straightforwardly since the Chern-Simons term is known to be
 gauge invariant up boundary terms, and the coupling term depends on $A$ only
 trough the field strength which is also gauge invariant.

In fact, considering the redefinitions $A_{\mu} = \Delta^{-1} {\tilde A}_{\mu} \Delta +
\Delta_{\mu}$ \, , $f_\mu = \Delta^{-1} {\tilde f}_\mu \Delta$, we get up to boundary terms,

\bea {\cal S}_{Parent}[A,f]&\equiv& {\cal S}_{CS}[{\tilde A }] - 
\int d^3x Tr \, \left( \epsilon^{\mu\nu\lambda}\,
 \Delta^{-1}~F_{\nu \lambda}[{\tilde A }]~\Delta~ f_{\mu} + m ~ f_{\mu} f^{\mu}  \right)
 \nonumber\\
&=& {\cal S}_{CS}[{\tilde A }] - \int d^3x Tr \left(\epsilon^{\mu\nu\lambda}\,
\Delta^{-1}F_{\nu \lambda}[{\tilde A }]{\tilde f}_{\mu}  \Delta  +
 m \,\Delta^{-1} {\tilde f}_{\mu} {\tilde f}^{\mu} \Delta^{-1} \right).\eea
Therefore, \be {\cal S}_{Parent}[A,f] \equiv {\cal S}_{Parent}[{\tilde A},{\tilde
f}].\ee
Varying ${\cal S}_{Parent}$ with respect to $f$, we obtain
 \be
f^{\mu a}=- {1 \over 2 m }\epsilon^{\mu\nu\lambda}\, F_{\nu\lambda}^{~~a}[A];
 \ee
plugging this back into (\ref{masterDJ}), and using \be
\epsilon^{\mu\nu\alpha}\epsilon_{\mu\nu\lambda} =
2\,\delta_\lambda^\alpha \label{ep} ,\ee
we recover
the YMCS-action, Eq. (\ref{510}).

Now, following strictly the standard program of the master action
approach \cite{suecos}, we must vary the parent action with respect to $A$,
and use the resulting equation to solve $A$ in terms of the other
field, $f$. Finally, one shall eliminate $A$ from the action.

Now, we vary with respect to $A$ and obtain:
 \be
2\epsilon^{\mu\nu\lambda}\,[
\partial_{[\nu}A_{\lambda]}^{~a} + \tau^{abc}\,
{A}_{\nu}^{~b}\,{A}_{\lambda}^{~c} -
 \partial_{[\nu}f_{\lambda]}^{~~a} - 2\tau^{abc}\,{A}_{\nu}^{~b} 
\,{f}_{\lambda}^{~c}]=0 ,\ee
by using (\ref{ep}), one can eliminate the Levi-Civita symbol, 
and from (\ref{391}) we can rewrite this equation as \be \label{eqf}
F_{\nu\lambda}^{~~a}[A_\mu -f_\mu ]= \tau^{abc} \,{f}_{\nu}^{~b} 
\,{f}_{\lambda}^{~c}  . \ee 
In the Abelian case this is \be F_{\nu\lambda}[A_\mu -f_\mu ]=0,\ee
then we have \be A_\mu =f_\mu  + \Delta_\mu . \ee Putting this back into the
action (\ref{masterDJ}) , we recover the SD theory (\ref{180}) up to
boundary terms.

The solution to the general equation (Non-Abelian) (\ref{eqf}) is
less understood; this is the origin for the difficulties for
establishing duality with the SD-model.

In the non-Abelian case, the field strength does not determine the gauge
potentials $(A_\mu -f_\mu )$ up to gauge transformations; this is known as
the Wu-Yang ambiguity \cite{WY}. In other words, the operator $F$ cannot
be inverted in equation (\ref{eqf}) in a unique way \footnote{In the
Fock-Schwinger gauge, the ( non-local ) solution of (\ref{eqf}) is
$ A_{\lambda}^{~a}=f_{\lambda}^{~a}  + \int_0^1 dt \, t \, x^{\nu} 
(\tau^{abc} \,{f}_{\nu}^{~b} 
\,{f}_{\lambda}^{~c})|_{t x^{\mu}}$, where $x^{\mu}$
is the space-time point \cite{FSh}. Notice that the non-local part, is second order in $f$.}.

  In Ref. \cite{KLRvN}, a solution is found by using the Fock-Schwinger
gauge, yielding a (second order in $f$) non-local solution.

We propose an alternative way to see this and tackle this problem.
Let us recall that one must find a {\it functional} solution, $A_\mu =A_\mu [f_\nu ]$ of this
equation and replace it into the action (\ref{masterDJ}), which
will result expressed in terms of $f$. However, one can assume
that a solution exists in this way at least perturbatively.

Let´s assume a
formal development of this functional in the form : \be
\label{dev} A_\mu = A_\mu^{(0)} + A_\mu^{(1)}[f_\nu ] + A_\mu^{(2)}[f_\nu ] +....\ee
 where $A_\mu ^{(0)}$ is independent of $f_\mu $, $A_\mu ^{(1)}$ is first order in $f_\mu $,
thus this shall be a linear functional (it may be a no-local
operator) of $f_\mu $, $A_\mu ^{(2)}$ is second order in $f_\mu $ and so on.

 Actually, we admit that the functional $A_\mu [f_\nu ]$ is analytical
enough in order to admit this
development (at least to first order). One can perform a perturbative
 analisys of the solutuion and solve this order by order.

Putting this development into equation (\ref{masterDJ}), and
assuming that this is satisfied to each order, we obtain two
equations for the zeroth and first orders respectively; the zeroth order 
is:

\be \epsilon^{\mu\nu\lambda}\,
F_{\nu\lambda} [A_\mu ^{(0)}]=0 .\ee Thus, using once more (\ref{ep}), this
reads

\be\label{ord0}  F_{\nu\lambda}[A_\mu ^{(0)}]=0 .\ee

 This implies
that the zeroth order corresponds to a pure gauge which does not
contribute to the action (\ref{masterDJ}). So, $A_\mu ^{(0)}$ can be dropped out of
the solution (\ref{dev}) \footnote{As it has been shown above, one can redefine 
$(A_{\mu}\,, f_\mu )\to (\Delta^{-1} {\tilde A}_{\mu} \Delta +
\Delta_{\mu} \, , \Delta^{-1} {\tilde f}_\mu \Delta )$, 
to obtain an equivalent parent action. Note also that these transformations do
 not change the order (in $f$) of the expressions.}.

The first order equation reads,
\be 
\label{ord1}  
\partial_{[\nu}\left( A^{(1)~a}_{\lambda]}-f^{~a}_{\lambda]}\right) + \tau^{abc}\,\left({
A}^{(1)~b}_{\nu}-f^{~b}\right)\,{A}^{(0)~c}_{\lambda}=0. \ee

Since we are interested in obtaining the self-dual model whose
action is third order in the potential field, let us substitute
the perturbative solution (\ref{dev}) into the master action and keep 
terms of third order in $f$,

\be {\cal S}={\cal S}_{Parent}[A^{(1)}, f]+ 
\epsilon^{\mu\nu\lambda}\, A^{(2)}_{\mu}\partial^{}_{[\nu}A^{(1)}_{\lambda]}+
 \epsilon^{\mu\nu\lambda}\,
A^{(1)}_{\mu} \partial^{}_{\nu}A^{(2)}_{\lambda} - 2\epsilon^{\mu\nu\lambda}\,
 f_{\mu} \partial^{}_{[\nu}A^{(2)}_{\lambda]} + o^4(f). \ee

Integrating out by parts, we obtain:

 \be \label{pertf}{\cal S}={\cal S}_{Parent}[A^{(1)},\, f] + \epsilon^{\mu\nu\lambda}\,
2[A^{(1)}_{\mu} -f_{\mu} ]\partial^{}_{[\nu}A^{(2)}_{\lambda]}+o^4(f) \ee

 Then, we can make a crucial observation in order to find the dual action: 
{\it only the first and second orders contribute to a dual third order
action}.

Let us calculate a solution for the first order. Below, we shall
prove that the second order will not be actually needed.

Like in the Abelian case, we can see that \be \label{sol} A_\mu ^{(1)}=f_\mu \, , \ee
and $A_\mu ^{(0)}=\Delta_\mu $ is a solution to (\ref{ord0}) and (\ref{ord1})
\footnote{Notice that (\ref{eqf}) is equivalent to $F(A-f)=0$
up to second order. This implies that the difference $A-f$, up to second order,
 is a pure gauge.
Thus, we may conclude that solution (\ref{sol}) is essentially unique.}.

Thus, using the fact discussed above, that a pure gauge is
irrelevant for the action, we can write 
\be
\label{sol1}
 A_\mu [f]= f_\mu  + A_\mu^{(2)}[f].
\ee

Substituting this into (\ref{pertf}), we can see that the second
term identically vanishes and a second order, therefore,
$A^{(2)}[f]$ will contribute to the action only in its {\it
fourth} order. Finally, we get:

\bea {\cal S}[f]&=&{\cal S}_{Parent}[ f_\mu  + A_\mu^{(2)}[f] \, , f_\mu ]+ o^4(f) \nonumber\\
&=&
 -\epsilon^{\mu\nu\lambda}\,
 \left[{f}_\mu^{~a}\partial^{}_{\nu}f_{\lambda}^{~a} +
 \,\frac{2\tau^{abc}}{3}\,{ f}_{\mu}^{~a} \,{ f}_{\nu}^{~b} 
\,{f}_{\lambda}^{~c}\right] - m \, f_{\mu}^{~a} f^{\mu a } +
o^4(f).
 \eea

We may rescale $f_\mu  \to {1 \over 2 } f_\mu $ and recover the SD-theory,

\be {\cal S}[f]= - {1 \over 4 }\epsilon^{\mu\nu\lambda}\,
 \left[{f}_\mu^{~a}\partial_{\nu}f_{\lambda}^{~a} +
\,\frac{\tau^{abc}}{3}\,{ f}_{\mu}^{~a} \,{ f}_{\nu}^{~b} 
\,{f}_{\lambda}^{~c}\right]  - ( { m \over 4 })f_{\mu}^{~a} f^{\mu a }  +
o^4(f).
 \ee

 This completes the proof of our main statement.

One may conclude that YMCS is (dual) equivalent to a
theory (described by the field $f$) which
coincides with the Self-Dual model for an
arbitrary coupling constant, $m^{-1}$, up to fourth order in $f$.
The well-known non-local contributions would appear at higher
orders in $f$. The perturbative approach in $f$ has been adopted as an artifact
 to solve the equation yielded by the parent action approach. Actually,
the parent action interpolates the two theories to this order
in $f$; and this is sufficient to ensure the equivalence between the models,
since the self-dual model is already third order in $f$.
 
This result shall have useful consequences for the bosonization
  identities
  between the massive
 Thirring model and the topologically massive model, whenever
  the fermions carry non-Abelian charges \cite{clI,FS,FS2}; we are 
presently addressing to this question and we shall report on it \cite{boson}.

Here, our strategy was somewhat different with respect
 to the usual analysis. We have tackled this problem from a perturbative point of view;
 which may be helpful
to solve similar problems and to establish other dual equivalences between models, besides
 the additional
 advantage of rendering more straightforward the treatment of non-Abelian mathematical
 structures. This is perhaps, the most relevant application of this work.

{\bf Aknowledgements}: The author is indebted to Prof. C. Wotzasek for pointing out the relevance
 of the problem and Prof. J. A. Helayel-Neto 
for invaluable discussions and pertinent corrections on the 
manuscript. Thanks are due to the GFT-UCP
 for the kind hospitality. CNPq is also acknowledged for the invaluable financial help.


\begin{thebibliography}{99}

\bibitem{DJ} S. Deser and R. Jackiw, Phys. Lett. B 139 (1984) 2366.

\bibitem{suecos} For a recent review in the use of the master action
 in proving duality in diverse areas see: S. E. Hjelmeland, U. Lindstr\"om, UIO-PHYS-97-03, May 1997.
e-Print Archive: hep-th/9705122.

\bibitem{KLRvN}A. Karlhed, U. Lindstr\"om, M. Ro\v{c}ek and P. van Nieuwenhuizen,
Phys. Lett. B 186 (1987) 96.
\bibitem{BFMS} N. Brali\'c, E. Fradkin, V. Manias and F. A. Schaposnik,
Nuc. Phys. B 446 (1995) 144.
\bibitem{TPvN} P. K. Townsend, K. Pilch and P. van Nieuwenhuizen, Phys.
Lett. B 136 (1984) 38.
\bibitem{BBG}N. Banerjee, R.Banerjee and S.Ghosh, Nucl. Phys. B 527 (1998) 402.


\bibitem{AINRW} D. Bazeia, A. Ilha,
 J.R.S. Nascimento, R.F. Ribeiro, C. Wotzasek Phys.Lett.B510:329-334, 2000
\bibitem{clI} A. Ilha, C. Wotzasek Nucl.Phys.B604:426-440, 2001.

\bibitem{clI2} M.A. Anacleto, A. Ilha, J.R.S. Nascimento, R.F. Ribeiro, C. Wotzasek
Phys.Lett.B504:268-274, 2001.
''Duality equivalence between nonlinear selfdual and topologically massive models",
A. Ilha, C. Wotzasek, hep-th/0106199.

\bibitem{annals} R.Jackiw, S. Deser and Templeton, Ann. Phys. 140 (1982) 372.
\bibitem{FS} E.~Fradkin and F.A.~Schaposnik, Phys.~Lett.  B 338 (1994) 253.
\bibitem{FS2} J.C. Le Guillou, E.F. Moreno, C. Nunez,
 F.A. Schaposnik, Mod. Phys. Lett. A12 (1997) 2707.

\bibitem{WY} T. T. Wu and C. N. Yang, Phys.Rev. D 12 (1965) 3843. 

\bibitem{FSh} V. A. Fock, Sov. Phys. 12 (1937) 404; J. Schwinger, Phys. Rev. 82 (1952) 684. 

\bibitem{boson}  M. Botta Cantcheff, J. A. Helayel-Neto , work in progress.
\end{thebibliography}
\end{document}